\documentclass[reprint,aps,amsmath,amssymb]{revtex4-1}

\usepackage{graphicx}
\usepackage{dcolumn}
\usepackage{bm}
\usepackage{hyperref}

\usepackage{amssymb,euscript,latexsym,amsmath}
\usepackage{graphicx}
\usepackage[dvips]{epsfig}
\usepackage{color}
\usepackage{epstopdf}
\usepackage{setspace}
\usepackage{subfigure}

\newcommand{\abs}[1]{\left| #1 \right|} 

\newcommand{\conj}[1]{\overline{#1}}
\newcommand{\real}[1]{\mathrm{Re} \left[ #1 \right]}

\newcommand{\eee}[1]{\mathrm{e}^{ #1 }}
\newcommand{\ii}{\mathrm{i}} 

\begin{document}

\preprint{APS/123-QED}
%
%
%
%
%
%
%
%
%


\title{Circularly-confined microswimmers  exhibit multiple global patterns}

\author{Alan Cheng Hou Tsang and  Eva Kanso} \thanks{corresponding author: kanso@usc.edu}
 \affiliation{Aerospace and Mechanical Engineering, University of Southern California, Los Angeles, CA 90089}

\begin{abstract}
Geometric confinement plays an important role in the dynamics of natural and synthetic microswimmers from bacterial cells to self-propelled particles in high-throughput microfluidic devices.  However, little is known about the effects of geometric confinement on the emergent global patterns in such self-propelled systems. Recent experiments on bacterial cells give conflicting reports
of cells spontaneously organizing into a spiral vortex in a thin cylindrical droplet and cells aggregating at the inner boundary in a spherical droplet. We investigate here, in an idealized physical model, the interplay between geometric confinement and level of flagellar activity on the emergent collective patterns. We show that decreasing flagellar activity induces a hydrodynamically-triggered transition in confined microswimmers from swirling to global circulation (vortex) to boundary aggregation and clustering. These results highlight that the complex interplay between confinement, flagellar activity and hydrodynamic flows in concentrated suspensions of microswimmers could lead to a plethora of global patterns that are difficult to predict from geometric consideration alone.

\begin{description}
\item[PACS numbers]
47.63.Gd,  87.18.Hf, 05.65.+b
\end{description}
\end{abstract}

\maketitle



The physics of natural and synthetic microswimmers is an active area of research both at the individual and collective levels. At the individual level, a great deal of attention is given to deciphering or devising mechanisms for self-propulsion in viscously dominant flows~\cite{lauga:rpp2009a, pak:arxiv2014a}. When acting together, these microswimmers  or ``units'' typically exhibit rich emergent dynamics  at the system's scale, which is often several orders of magnitude larger than the scale of the individual unit~\cite{mendelson:jbact1999a,dombrowski:prl2004a,cisneros:ef2007a,sokolov:prl2007a,sokolov:pre2009a,dunkel:prl2013a}. 
Several computational models have been proposed to explain the mechanisms, long-range hydrodynamic versus steric local interactions, responsible for these complex emergent dynamics. Most models have primarily focused on populations of unconfined microswimmers (unbounded three-dimensional domains) or subject to weak confinement~\cite{hernandez-ortiz:prl2005a,hernandez-ortiz:jpcm2009a,saintillan:prl2007a,saintillan:prl2008a,saintillan:pof2008a,saintillan:jrsi2012a,ishikawa:jfm2008a,ishikawa:prl2008a,evans:pof2011a,lushi:cs2013a}. These models successfully capture the emergence of large scale flows and collective coherent structures observed in many experiments on bacterial suspensions~\cite{mendelson:jbact1999a,dombrowski:prl2004a,cisneros:ef2007a,sokolov:prl2007a,sokolov:pre2009a,dunkel:prl2013a}.  However,  in many experimental set-ups as well as in their natural environment, bacteria are often subject to different types of boundary confinement. In fact, confinement of micro-swimmers, natural and synthetic, is of great interest in many applications of science and engineering, from bacterial biofilms to micro-fluidics devices~\cite{miller:armb2001a,edmunds:jbr2013a,bricard:n2013a}. 

The effects of geometric confinement on the dynamics of individual swimmers  and swimmer populations are becoming the center of attention of several experimental and theoretical studies. 
Recent experiments on confined bacterial suspensions within a cylindrical droplet show that, if the radius of confinement is below a critical value, the bacteria can spontaneously organize themselves into a spiral vortex encircled by a counter-rotating cell boundary layer~\cite{wioland:prl2013a}.
This remarkable self-organization is attributed to the interplay between hydrodynamic and boundary interactions~\cite{lushi:pnas2014a}. Other experiments on confined bacterial suspensions in a spherical water droplet report that the bacteria, at relatively small concentrations, tend to aggregate at the inner droplet boundary rather than forming a global vortex~\cite{vladescu:arxiv2014a}. In another context of geometric confinement where swimmers are strongly confined in a two-dimensional \emph{Hele-Shaw} set-up, continuum theoretical models predict interesting instabilities in the emergent collective behavior that are qualitatively distinct from those observed in unconfined swimmers~\cite{brotto:prl2013a}. Meanwhile discrete simulations show new forms of collective motions, including swirling, orientational order and aggregation~\cite{lefauve:pre2014a, tsang:pre2014a}.

\begin{figure*}[!t]
\centerline{\includegraphics[width=0.6\textwidth]{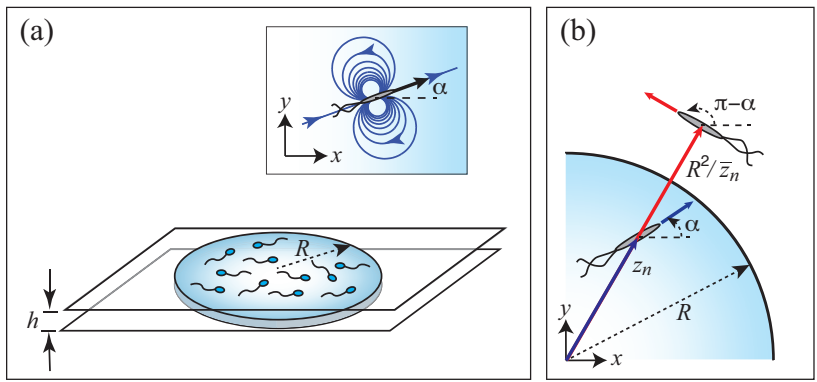}}
\caption[]{Microswimmers  in Hele-Shaw and circular confinement: (a) schematic of the set-up and the  dipolar far-field flows created by a swimmer in Hele-Shaw confinement (inset); (b) Effect of the circular confinement is accounted for using an image system outside the circular domain as dictated by the Milne-Thompson theorem.}
	\label{fig:schematic}
\end{figure*}  

One of our main goals in this work is to understand how the interplay of both \emph{Hele-Shaw} and \emph{circular confinement} affects the emergent collective dynamics in the context of an idealized model that takes into account the long-range hydrodynamic interactions among microswimmers. 
Microswimmers in Hele-Shaw geometries have a distinct hydrodynamic signature in the sense that the far-field flow is that of a two-dimensional (2D) potential source dipole as opposed to the three-dimensional (3D) force dipole in the unbounded case~\cite{brotto:prl2013a}. The dipolar far field is independent of the transport mechanism, be it driven particles or self-propelled swimmers, pushers or pullers~\cite{desreumaux:epje2012a,beatus:pr2012a}. It is rooted in the fact that the basic physics is that of a Hele-Shaw potential flow~\cite{beatus:pr2012a}. Further, due to friction with the nearby walls, confined micro-swimmers with head-tail polarity, due to either geometric properties~\cite{brotto:prl2013a} as in self-propelled colloids or Janus particles~\cite{walther:cr2013a,wensink:pre2014a} or due to flagellar activity~\cite{tsang:pre2014a} as in bacterial cells, reorient in response to both the local flow field and its gradient. 

For concreteness, we briefly review the model presented  in~\cite{tsang:pre2014a}. The dynamics of a population of $N$  microswimmers in Hele-Shaw confinement can be expressed in concise complex notation as
\begin{equation}
\begin{split}
\label{eq:formulation:eom}
  \dot{\conj{z}}_n  & = U \eee{-\ii \alpha_n}+ \mu \conj{w}(z_n)+ V_n, \\[1ex]
  \dot{\alpha}_n  & =  \real{\nu \conj{w} \ii  \eee{\ii \alpha_n}+\nu^{\prime} \frac{d \conj{w}}{dz} \ii \eee{2 \ii \alpha_n}}.
  \end{split}
\end{equation}
Here, $z_n$ denotes the swimmer's position in the complex plane $z = x+\ii y$ $(\ii = \sqrt{-1})$ and $\alpha_n$ its orientation, $n=1,\ldots, N$. The operator Re$[\cdot]$ takes the real part of the expression in bracket whereas the overline $\conj{(\cdot)}$ denotes the complex conjugate of the expression in parenthesis. A local replusive velocity $V_n$ is introduced to ensure no collision among swimmers. Each swimmer has a self-propelled speed $U$ and is advected by the local flow velocity $w$. It is understood that all variables and parameters are non-dimensionalized based on characteristic length and time scales dictated by the individual swimmers. 
The  dimensionless parameters $\mu$, $\nu$ and $\nu^\prime$ are translational and rotational mobility coefficients whose values depend on the geometric properties of the swimmer and its flagellar activity. More specifically,  $\mu$ arises from the balance of drag and friction acting on each swimmer due to the \emph{Hele-Shaw} confinement and is in the range $0< \mu<1$.
The value of $\nu$ reflects the degree of flagellar activity (or geometric polarity) and could decrease from a positive value $(\nu > 0)$ for vigorously-beating flagella (or large-tail Janus particles) to a negative value $(\nu < 0)$ for weakly-beating flagella (or large-head Janus particles),
thus causing the swimmer to switch from reorienting with the local flow velocity to reorienting in the  direction opposite to the local flow. Besides reorientation in response to the local flow, swimmers can also reorient in response to the local flow gradient as reflected by the $\nu^{\prime}$ term, which is consistent with the classical Jeffery's orbit~\cite{jeffery:prsa1922a}. The effect of $\nu^{\prime}$ term is small when the absolute value of $\nu$ is sufficiently greater than zero as shown in~\cite{tsang:pre2014a}. Thus, for simplicity, we set  $\nu^\prime$ to be identically zero ($\nu^{\prime}=0$) in the present work.

To close the model in~\eqref{eq:formulation:eom}, we need to evaluate the velocity field $w(z)$ induced by $N$ microswimmers, each creating a far-field flow equivalent to that of a source dipole which decays as $1/|z|^2$. More specifically, the conjugate velocity created by a source dipole located at $z_n$ at an orientation $\alpha_n$ is given by $\sigma e^{\ii\alpha_n}/(z-z_n)^2$, where $\sigma$ is the dipole strength. In~\cite{tsang:pre2014a},  we considered $N$ microswimmers in a doubly-periodic domain and presented a closed-form solution for the velocity field $w$ created by this doubly periodic system in terms of the Weierstrass elliptic function. One of the main contributions of the present manuscript lies in deriving a closed-form expression for the velocity field created by $N$ such swimmers in a circularly-confined domain. We then investigate numerically the emergent behavior afforded by a population of microswimmers subject to  Hele-Shaw and circular confinement as a function of the confinement radius and flagellar activity.


To take into account the effect of circular confinement on the hydrodynamic signature of the microswimmers,  we apply the \emph{Milne-Thomson circle theorem}, see, e.g.,~\cite{batchelor:1967a}. This theorem states that for any potential flow with an associated complex potential function $F(z)$, analytic in the considered domain, a circular boundary of radius $R$ can be constructed at the origin using the modified complex potential
\begin{equation}
\label{eq:circlethm}
	G(z)=F(z)+\conj{F(R^2/\conj{z})}.
\end{equation}
The second term in the expression enforces the imaginary part of the modified complex potential function $G(z)$ to be zero at $\abs{z}=R$, thus creating a circular boundary of radius $R$.

\begin{figure*}[!t]
\centerline{\includegraphics[width=0.75\textwidth]{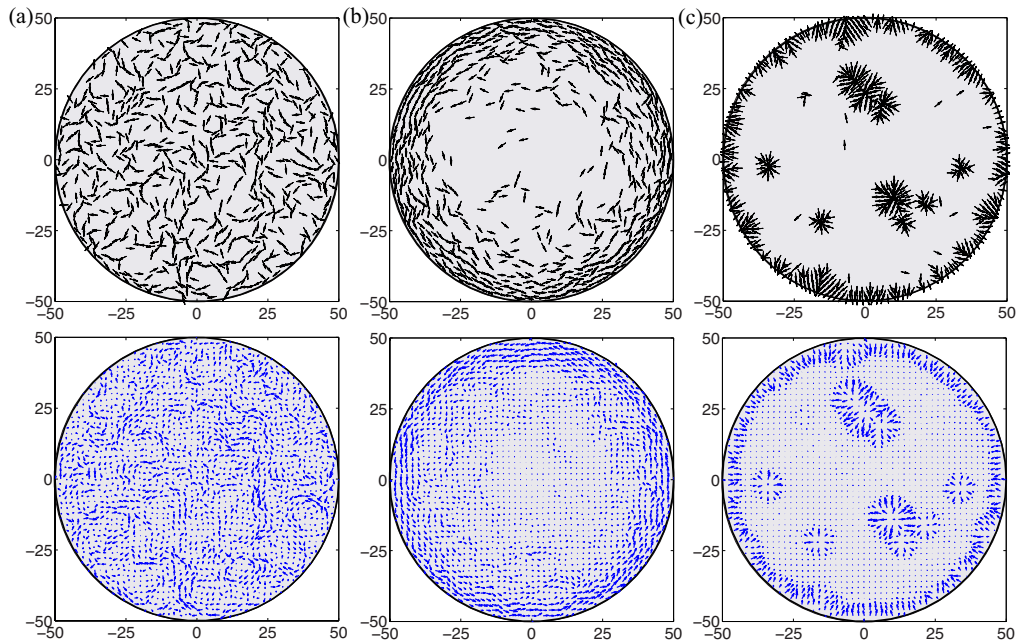}}
\caption[]{Snapshots for the emergence of three global modes: (a) chaotic swirling ($\nu=3$); (b) stable circulation ($\nu=0.5$); (c) boundary aggregation ($\nu=-0.5$). The corresponding flow fields are depicted at the bottom row. For all three simulations, $N=750$ and $R=50$.}
	\label{fig:behavior}
\end{figure*}

In \emph{Hele-Shaw} confinement, the leading-order flow disturbance created by each swimmer is that of a potential dipole, see Fig.~\ref{fig:schematic}(a) and~\cite{brotto:prl2013a,lefauve:pre2014a, tsang:pre2014a}. The complex potential function for a system of $N$ swimmers is therefore given by
\begin{equation}
\label{eq:dipolepotential}
	  F(z) =\sum_{j=1}^N -\frac{\sigma \eee{\ii \alpha_j}}{z-z_j},
\end{equation}
where $\sigma= a^2 U$ denotes the dipole strength, with $a$ being the effective radius of the swimmer. The complex conjugate velocity at each swimmer's position, excluding the self-induced flow disturbance, is given by
\begin{equation}
\label{eq:velocitycirc}
	\conj{w}(z_n)=\left[\left. \frac{dG}{dz}\right|_{z=z_n}-\frac{\sigma \eee{\ii \alpha_n}}{(z-z_n)^2}\right].
\end{equation}
By virtue of \eqref{eq:circlethm} and \eqref{eq:dipolepotential}, Eq.~\eqref{eq:velocitycirc} becomes
\begin{equation}
\label{eq:velocity}
	\conj{w}(z_n)=\sum_{j \neq n}^N \frac{\sigma \eee{\ii \alpha_j}}{(z_n-z_j)^2}+ \sum_{j =1}^N \frac{\sigma (R^2/\conj{z}_j^2) \eee{\ii(\pi- \alpha_j)}}{(z_n-R^2/ \conj{z}_j)^2}.
\end{equation}
That is, constructing the flow field created by  a swimmer located at $z_n$ and orientation of $\alpha_n$ in a circular boundary of radius $R$ amounts to placing an imagine swimmer at location $R^2/\conj{z}_n$, with orientation $\pi-\alpha_n$ and dipole strength $\sigma(R^2/\conj{z}_n^2)$. A schematic of the image system is illustrated in Fig.~\ref{fig:schematic}(b). 

We introduce a collision avoidance mechanism among swimmers and between a swimmer and the circular boundary based on the rapidly-decaying repulsive part of the Leonard- Jones potential
\begin{equation}
	\label{eq:replusion}
	V_n  =  \sum_{j \neq n}^N f_0 \left[ \left(\frac{2a}{\abs{z_n-z_j}}\right)^{13}\frac{\conj{z}_n-\conj{z}_j}{\abs{z_n-z_j}} +  \left(\frac{a}{\abs{d}}\right)^{13}\frac{\conj{z}_n-\conj{d}}{\abs{z_n-d}} \right],
\end{equation}
Here, $f_0$ is a scaling parameter that characterizes the repulsion strength, $d$ is the shortest vector joining the swimmers and the circular wall (parallel to the perpendicular bisector of the circle). The first term ensures collision avoidance among swimmers whereas the second term guarantees that no swimmer penetrates the circular boundary. These near-field interactions decay rapidly outside a small excluded area centered around $z_n$. Their rapid decay ensures that the order of the far-field hydrodynamic interactions is preserved. Equations \eqref{eq:formulation:eom}, \eqref{eq:velocity} and \eqref{eq:replusion} form a closed system of equations for $N$ confined swimmers in a circular domain of radius $R$ that take into account the swimmer-generated flows and level of flagellar activity.
%
%



We examine the emergent global modes of populations of $N$ swimmers as a function of the confinement radius $R$ and flagellar activity $\nu$. We normalize $U$ and $\sigma$ to 1 and set $\mu=0.5$. We fix the area fraction of the swimmers $\beta=Na^2/R^2$ to 0.3, that is, we vary the number of swimmers $N$ in the domain as $R$ changes. We focus on initial conditions for which the swimmers are initially randomly oriented but spatially homogeneous. We perform Monte-Carlo type simulations in the sense that for each set of parameters, we run multiple trials corresponding to different sets of initial conditions, each set taken from a uniform probability distribution function. 

We observe that, as we vary the level of flagellar activity $\nu$ for fixed radius $R$, the swimmers organize into three distinct types of global modes: chaotic swirling, stable circulation and boundary aggregation. Snapshots for representative simulations are shown in Fig.~\ref{fig:behavior}, together with their corresponding instantaneous velocity fields. The emergence of chaotic swirling mode 
was observed in our previous study on confined swimmers in a doubly-periodic domain~\cite{tsang:pre2014a}, and was explained on the ground that swimmers have vigorously-beating flagella ($\nu$ positive) tend to tail-gate other swimmers ahead and form chain-like structures. This results in vortices and swirls in the velocity field, see Fig.~\ref{fig:behavior}(a) left, that is reminiscent of the bacterial turbulence observed in numerous experiments~\cite{mendelson:jbact1999a,dombrowski:prl2004a,cisneros:ef2007a,sokolov:prl2007a,sokolov:pre2009a,dunkel:prl2013a}. 

A surprising emergent mode arises when we reduce $\nu$ to smaller positive values. In this case, the swimmers spontaneously organize into stable circulation, characterized by a single vortex with swimmers concentrating at the circular boundary, see Fig.~\ref{fig:behavior}(b). The circulation occurs with equal probability in clockwise and anti-clockwise direction. Similar phenomenon was observed in the recent experiments and simulations of bacterial suspensions in confined circular domains,~\cite{wioland:prl2013a, lushi:pnas2014a}. However, in contrast to the single vortex state presented here, \cite{wioland:prl2013a, lushi:pnas2014a} report a pair of counter-rotating vortices, constituting a single vortex of swimmers encircled by a counter-rotating layer of swimmers.

As $\nu$ decreases to a negative value, indicating swimmers with even weaker flagellar beating, the swimmers aggregate at the circular boundary or form clusters inside the domain, see Fig.~\ref{fig:behavior}(c). 
The formation of `inner' clusters
was observed in our previous study on confined swimmers in a doubly-periodic domain~\cite{tsang:pre2014a}. It was explained based on the fact that swimmers with weakly-beating flagella tend to reorient in the opposite direction to the local flow created by nearby swimmers and, thus, tend to aggregate and cluster. These clusters have the shape of an asterisk and create a source-like convective flow that attracts other swimmers to the cluster, potentially causing it to develop into large cluster. The aggregation at the circular boundary can be attributed to a similar phenomenon where the swimmers tend to cluster with their image system outside the circular domain. To our best knowledge, aggregation and clustering at the circular boundary has not been reported in theoretical models before but has been observed in the recent experimental work of~\cite{vladescu:arxiv2014a} albeit for bacteria in 3D spherical confinement as opposed to the 2D circular confinement considered here.

\begin{figure}[!t]
\centerline{\includegraphics[width=0.5\textwidth]{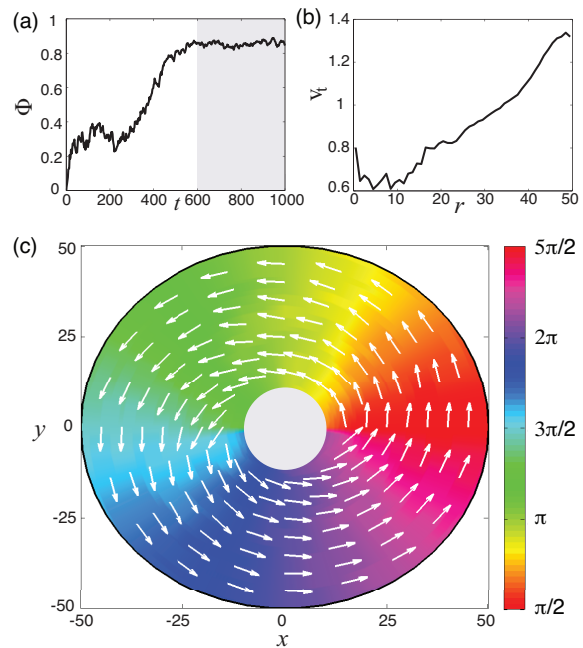}}
\caption[]{Statistical data for the stable circulation mode shown in Fig.~\ref{fig:behavior}b: (a) vortex order parameter $\Phi$; (b) mean tangential speed; (c) mean polarization field. The mean polarization field and tangential speed are computed based on the averaging of data over the shaded time range, which a stable vortex order has been developed. }
	\label{fig:polar}
\end{figure}

We use a number of statistical measures to assess the observed global structures.  
We focus on the statistics of the circulation mode. The statistics of the other two modes (swirling and aggregation) have been described in details in our previous study~\cite{tsang:pre2014a} and are omitted here for brevity. Following~\cite{wioland:prl2013a,lushi:pnas2014a}, we introduce the vortex-order parameter
\begin{equation}
\label{eq:vortexorder}
	\Phi=\frac{\sum_j \text{v}_{t,j}/\sum_j \abs{\dot{z}_j}-2/\pi}{1-2/\pi},
\end{equation}
where $\text{v}_{t,j}=\abs{\dot{z}_j\conj{t}_j+\dot{\conj{z}}_j t_j}/2$ is the tangential speed of each swimmer and $t_j$ is the vector parallel to the tangent of the circle at the location of the swimmer. one gets $\Phi=1$ for steady circulation and $\Phi=0$ for disordered state. 
The change in $\Phi$ with time is presented in Fig.~\ref{fig:polar}(a) for the case shown in Fig.~\ref{fig:behavior}(b). The value of  $\Phi$ levels off when the swimmers develop into a stable circulation. 
Fig.~\ref{fig:polar} also depicts the  mean tangential speed and mean polarization field  for the same case.  We compute the mean tangential speed of the swimmers as a function of the distance $r=\sqrt{x^2+y^2}$ from the origin, see Fig.~\ref{fig:polar}(b). The tangential speed increases with $r$, indicating that the swimmers have a higher vortex order closer to the circular boundary. 
The polarization field shows the average orientation of the swimmers at different location in the vortex state and is computed by interpolating the orientation of the swimmers to a polar grid. The center of the polarization field, where no clear orientation can be determined, is left out in gray color. The formation of vortex structure can be seen clearly from the polarization field as shown in Fig.~\ref{fig:polar}(c).

\begin{figure}[!t]
\begin{center}
\includegraphics[width=0.475\textwidth]{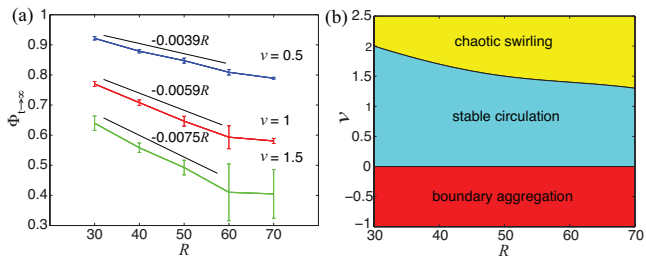}
\end{center}
\caption[]{(a) The time-averaged mean  and standard deviation of long-time developed $\Phi$ over various initial conditions. The standard deviations are denoted by the errorbars. (b) Phase diagram showing the three different global modes: chaotic swirling, stable circulation and boundary aggregation, as a function of $\nu$ and $R$.}
	\label{fig:phasediag}
\end{figure}

We next examine the effect of the interplay between flagellar activity $\nu$ and confinement radius $R$  on the developed circulation as measured by the vortex-order parameter $\Phi$. Fig. \ref{fig:phasediag}(a) shows the mean long-time $\Phi$ for several initial conditions and for different values of $\nu$ and $R$. As $R$ increases, the value of $\Phi$ decreases almost linearly when $R<60$ and the rate of decrease in $\Phi$ increases with $\nu$. Large error bars are observed in the cases of $R=60$ and $R=70$ for $\nu=1.5$, which mark the presence of an instability that prevents the swimmers from forming stable circulation. This indicates that the swimmers can only form stable circulation when the radius of the circular boundary is below a critical value whose precise value depends on the flagella activity.  The existence of critical radius for stable circulation is also observed in the aforementioned experiments of bacterial suspensions within a circular droplet~\cite{wioland:prl2013a, lushi:pnas2014a}. In their case, the value of $\Phi$ first increases with $R$, then levels to a constant value for a range of $R$ before $R$ exceeds a  critical value for which $\Phi$ suddenly drops. In our case, $\Phi$ is monotonically decreasing as a function of $R$ before $R$ reaches a critical value for which no stable circulation is observed. 
We set threshold value of $\Phi_{t \rightarrow \infty} = 0.5$ to distinguish between the stable circulation and chaotic swirling modes. We then map three global modes: chaotic swirling, stable circulation and boundary aggregation onto the 2D parameter space ($\nu$, $R$)
 depicted in Fig.~\ref{fig:phasediag}(b). This plot summarizes the main findings of this work and identifies the critical values of circular confinement $R$ that mark the transition from chaotic swirling to stable circulation for different values of flagellar activity $\nu$. 


Our minimal model  for microswimmers in Hele-Shaw and circular confinement accounts for  swimmer-fluid interactions, swimmer-generated flows, and level of flagellar activity and captures qualitatively the dynamics seen in experiments on confined bacterial suspensions. In particular, we showed the following:

(i) For vigorously-beating flagella, the global swirling pattern is reminiscent to turbulence-like patterns observed in many experiments on bacterial suspensions.

(ii) As the flagellar activity or the confinement radius decreases, the microswimmers stabilize into a global vortex with swimmers' concentration increasing towards the circular boundary.

(iii) Our model does not capture the counter-rotating layer of swimmers near the circular boundary reported experimentally and computationally in~\cite{wioland:prl2013a, lushi:pnas2014a}. 
This discrepancy can be attributed to two main differences between the two models. First, the leading order flow disturbance created by the swimmers are different in the two models: here, swimmers create a potential dipolar field that decays with distance from the swimmer as $1/r^2$. In~\cite{lushi:pnas2014a},  solving Stokes flow in 2D domains, a swimmer's dipolar flow decays as $1/r$. 
Further, here swimmers are subject to friction from the upper/lower walls while these effects are neglected in~\cite{lushi:pnas2014a}. Second,  steric interactions are added here to avoid direct collision but the orientation response in~equation \eqref{eq:formulation:eom} is purely hydrodynamic. In~\cite{lushi:pnas2014a}, steric interactions in the form of Gay-Berne potential were used  to account for both  collision avoidance and orientation of the swimmers at the circular boundary. Despite those discrepancies, the simpler model presented here is able to capture the leading-order global behavior which is that of stable circulation. 

(iv) As the flagellar activity decreases further so that the swimmers begin to reorient in the opposite direction to the local flow, the swimmers spontaneously cluster and aggregate at the circular boundary. This is an interesting global phenomenon that is reminiscent to the recently reported experimental results on bacterial cells aggregating at the boundary of a spherical droplet~\cite{vladescu:arxiv2014a}. 

These results lend a note of caution when drawing conclusions on emergent patterns, especially in experiments on living organisms and bacterial suspensions.  These experiments are often conducted under specific environmental conditions and parameter values. It is therefore important to emphasize that the observed global structures  may only be valid under those conditions. For example, while one may conclude based on~\cite{wioland:prl2013a, lushi:pnas2014a} that suitably designed boundaries provide a means for stabilizing and controlling order in active microbial systems, this conclusion is somewhat overreaching given that time-dependent bacterial properties, such as level of flagellar activity, may influence and alter the desired global order.

We thank E. Lushi for useful discussions.

\bibliography{reference}

\end{document}